\documentclass[doublecol]{epl2}
\usepackage{graphicx}
\usepackage{amsfonts}

\newcommand{\revision}[1]{{\bf #1}}

\newcommand{\chs}[1]{\hat{\chi}_s^{#1}}

\newcommand{\la}{\langle}
\newcommand{\ra}{\rangle}

\newcommand{\IR}{\mathbb{R}}

\title{Peeping at chaos: Nondestructive monitoring of chaotic systems by measuring
long-time escape rates}
\shorttitle{Peeping at chaos}
\author{L. A. Bunimovich\inst{1} \and C. P. Dettmann\inst{2}}
\shortauthor{L. A. Bunimovich \etal}
\institute{\inst{1} Applied \& Biological Contemporary Mathematics Program, Georgia Institute of Technology, Atlanta GA 30332-1060, USA\\
\inst{2} School of Mathematics, University of Bristol, Bristol BS8 1TW, UK}
\date{\today}
\abstract{
One or more small holes provide non-destructive windows to observe
corresponding closed systems, for example by measuring long time escape
rates of particles as a function of hole sizes and positions.  To leading
order the escape rate of chaotic systems is proportional to the hole size
and independent of position.
Here we give exact formulas for the subsequent terms, as sums of
correlation functions; these depend on hole size and position, hence
yield information on the closed system dynamics.  Conversely, the theory
can be readily applied to experimental design,
for example to control escape rates.}
\begin{document}
\maketitle

A fundamental issue in physical measurement (whether classical or quantum)
is ensuring that the system
to be studied is little affected by the observation.  This is clearly much
easier to satisfy if the measuring device lies outside the system.
A truly isolated system cannot affect its surroundings, and is impossible to
observe.  However it can be to a good approximation unaffected by a very
small hole, through which particles or radiation can carry information
about the internal dynamics.  Notice that theory and experiment often
work in opposite directions: theorists use closed systems to understand
open ones, while experimentalists do the reverse.

The initial motivation for this work came from atom-optics
billiards~\cite{MHCR01,FKCD01} in which escape properties were used to
distinguish regular and chaotic behavior, but for which a detailed theory
was absent, a deficit which the current work aims to address.
Our theory is however far more general than these experiments suggest;
billiards in which a particle moves in straight lines making specular
reflections with a boundary are relevant to any system with particles
in a relatively homogeneous cavity.  Experimental realizations have
included microwaves~\cite{S91,SS92,Gea92,PRSB00},
visible light~\cite{SZOL01}, phonons in quartz blocks~\cite{EGLNO96}, 
and electrons in semiconductors~\cite{Sea91,B99}.  Furthermore,
transport through small holes (in phase space) has applications as diverse as
the transition state theory of chemical reactions~\cite{WBW05a},
the migration of asteroids~\cite{WBW05b},
\revision{and passive advection in fluids~\cite{TSPT}.}  It can also help to
characterise chaos in relativity~\cite{ML01} and in Hamiltonian systems in
general~\cite{STN02}.  Finally, the ``hole'' can be a desired region in phase
space, for example a set from which we can subsequently control the dynamics
with small perturbations; the escape rate then gives information on how
long we may need to wait before attempting control~\cite{BP01}.
We emphasize that while our numerical example is a billiard, the method
is readily applicable to general volume preserving (for example Hamiltonian) systems. 

Just as black body radiation emanating from a cavity through a small hole
gives only a single piece of information (the temperature), we would expect
the transport properties of orbits in the phase space of open
systems to give only simple geometrical information to leading order.
The transit time (for orbits entering through a hole then exiting) is exactly given by
the ratio of hole to system sizes~\cite{Mei97}.  The escape rate (for orbits
initially in the system) is hard to characterize in general~\cite{DY06},
but is also often assumed to follow a similar equation~\cite{Mei97}.
In a previous Letter~\cite{BD05}, we showed a result of this type for the
regular circle billiard, with corrections computed analytically.
\revision{Stadium and related billiards with interior holes have also
been studied recently~\cite{NKLSW}.}
Here we consider general \revision{strongly} chaotic systems
\revision{in the limit of small holes,} and find an analytic expansion of
the escape rate, in which the ratio of hole and system sizes gives the
leading term, and corrections are given by correlation functions. Measurement
of the escape rate of a hole as a function of size and position, or of
two holes compared to the individual escape rates, then provides a window
through which to study the original (closed) dynamics.

The use of correlation functions to compute transport coefficients is
familiar from Green-Kubo formulas~\cite{EM90}.  Also, the escape rate from rather
specific systems of large spatial extent can be used to compute
general transport coefficients~\cite{GN90}.  Thus the appearance of
correlation functions in more general escape rate calculations should
not come as a surprise.
We also note that the periodic orbit formalism for computation of
escape rates of open chaotic systems~\cite{KT84,AAC90} is unsuitable for
the limit of small holes, as a very large number of periodic orbits would
be required for the hole(s) to be sufficiently covered. 

Our formalism is in the setting of volume preserving maps $\Phi:M\to M$,
such as Hamiltonian evolution considered at equally spaced times
(a ``stroboscopic map'') or when a certain condition is met
(a ``Poincar\'e map'').  We use $\langle f\rangle$ to denote
an integral of a phase variable $f:M\to \IR$ over the normalised
volume element of the phase space $M$.  \revision{Even when the dynamics
does not preserve the usual volume (for example leading to a strange
attractor), there is often a well defined fractal measure which is
preserved, albeit making calculations of correlation functions more
complicated.}

A particular case of a Poincar\'e map is that of a billiard system of a
particle moving uniformly
between specular collisions with the boundary $\partial D$ of a two dimensional
domain $D$, where $\Phi$ denotes the evolution from one collision to the next.
In this case the phase space volume element of the flow projects to the area
element $dl dp_{\parallel}$ of $M$ preserved under $\Phi$, where $l$ measures
arc length along the boundary and $p_{\parallel}$ is the component of
momentum parallel to the boundary after the collision.
Thus for a phase function $f$ on the billiard boundary we have
\begin{equation}
\langle f\rangle=\frac{\int f(l,p_\parallel) dl dp_{\parallel}}
{\int dl dp_{\parallel}}
\end{equation}
where the denominator is simply $2|{\bf p}||\partial D|$.

We now return to the general formalism and define some phase
variables on $M$.  Let the function $T:M\to \IR$ denote the time
from one collision to the next; this is necessary to relate
collision number with real time $t$.  We assume this function is
bounded from above. The holes are represented by a characteristic function
$\chi:M\to \IR$ equal to zero on a hole and one elsewhere; in the billiard
example this is normally a function only of $l$ and not $p_{\parallel}$.
Note that our approach here is to always use the dynamics of the closed system,
but ``kill'' escaping trajectories using a multiplication by $\chi$.
$\langle\chi\rangle\equiv 1-h$ is simply the fraction of $M$ not covered by
holes and $h$ is the size of the hole (relative to $|M|$).  Another
useful set of averages is $\tau_k\equiv\la T^k\chi\ra/\la\chi\ra$, the average
of the $k$th power of the collision time over the non-hole part of the system.
We will need a weighted characteristic function $\chi_s:M\to\IR$ defined by
$\chi_s\equiv e^{sT}\chi$, so that $\chi_0=\chi$.  It is
also useful to define phase functions with averages subtracted;
we find that the best way to do this is to set
$\hat{\chi}_s\equiv \chi_s/\langle\chi_s\rangle-1$ so that
$\hat{\chi}=\hat{\chi}_0=\chi/\langle\chi\rangle-1$.  We also define
$\hat{T}\equiv T-\tau_1$ and $\hat{\tau}_k=\la \hat{T}^k\chi\ra/\la\chi\ra$.

Assume that the phase space is filled with initial conditions (or mutually non-interacting particles)
of a uniform density with respect to the preserved volume element, then consider the survival probability
$P(t)$ for time $t$, so $P(0)=1$.  The exponential escape rate $\gamma$, \revision{defined for strongly chaotic
systems,} is
\begin{equation}
\gamma=-\lim_{t\to\infty}\frac{\ln P(t)}{t}
\end{equation}
which is well behaved in strongly chaotic systems
and is equal to the leading pole (that is, of smallest $s>0$) of the function
\begin{equation}
\int_0^\infty e^{st}P(t)dt=
\int_0^\infty e^{st}\langle\chi^0\chi^1\ldots\chi^{N_t}\rangle dt
\end{equation}
where $N_t$ is the number of collisions before time $t$ and a superscript
on a phase variable from now on will usually denote discrete time, ie
$\chi^j\equiv \chi\circ\Phi\circ\cdots\circ\Phi$, with the map $\Phi$ composed $j$ times.
We can rewrite this as
\begin{equation}
\int_0^\infty \langle\chi^0_s\chi^1_s\ldots\chi^{N_t}_s
\rangle dt=\sum_{N=0}^\infty\langle \chi^0_s\chi^1_s\ldots\chi^N_sT^N\rangle
\end{equation}
where $\chi_s$ is defined above, and we have ignored a factor bounded in
magnitude which does not affect the leading pole.
Now assuming that the long time escape is not dominated by orbits
with very small $T$ (hence $T^N$ is also effectively bounded),
the escape rate is given by the leading pole of
\begin{equation}
G(s)=\sum_{N=0}^{\infty}G_N(s)=\sum_{N=0}^{\infty}\langle \chi^0_s\chi^1_s\ldots\chi^N_s\rangle
\end{equation}
Substituting $\chi_s=\la\chi_s\ra(1+\hat{\chi}_s)$ to isolate the leading part,
and then expanding and grouping into terms with different numbers of $\hat{\chi}_s$, we find
\begin{equation}
G(s)=\sum_{N=0}^{\infty}\langle\chi_s\rangle^{N+1}\Gamma_N(s)
\end{equation}
where
\begin{eqnarray}
\Gamma_N(s)&=&\sum_{n=0}^{N+1}\sum_{0\leq j_1<j_2\ldots<j_n\leq N}
\langle\chs{j_1}\chs{j_2}\ldots\chs{j_n}\rangle\\
&=&1+\sum_{0\leq j<k\leq N}\langle\chs{j}\chs{k}\rangle
+\sum_{0\leq j<k<l\leq N}\langle\chs{j}\chs{k}\chs{l}\rangle
+\ldots\nonumber
\end{eqnarray}
Remarks:
\begin{enumerate}
\item The single sum ($n=1$) is absent since the average of $\hat{\chi}_s$
is zero by definition.
\item This expansion is divergent since the number of terms in each sum
gets larger with each term, and the higher order correlations are small
but finite.
\end{enumerate}

The sum for $G(s)$ diverges when the ratio of subsequent terms exceeds unity in the
limit, so at the first zero of
\begin{equation}
g(s)\equiv\lim_{N\to\infty} g_N(s)
\end{equation}
where
\begin{equation}
g_N(s)=\ln\frac{G_{N+1}(s)}{G_N(s)}
=\ln\langle\chi_s\rangle+\ln\Gamma_{N+1}(s)-\ln\Gamma_N(s)
\end{equation}
Expanding the logarithms and taking the limit, we find
\begin{equation}
g(s)=\ln\langle\chi_s\rangle+\sum_{n=2}^\infty Q_n(s)
\end{equation}
where the cumulants $Q_n(s)$ are
\begin{eqnarray}
Q_2(s)&=&\sum_{0<j}\langle\chs{0}\chs{j}\rangle\nonumber\\
Q_3(s)&=&\sum_{0<j<k}\langle\chs{0}\chs{j}\chs{k}\rangle
\end{eqnarray}
The higher cumulants are more complicated, for example $Q_4(s)$ includes
4-time correlations and products of 2-time correlations.
We expect this cumulant expansion (by analogy with~\cite{Det03}) to be
well defined if the system has multiple correlation functions decaying
faster than any power (which, a posteriori, is what we mean by ``chaotic'').
Decay with a power greater than unity will permit $Q_2(s)$ to exist, and
hence allow the second order formulas below to be used.

We are interested in the limit of small holes, ie small escape rate.  Thus
we expand these quantities in powers of $s$:
\begin{eqnarray}
\hat{\chi}_s&=&\hat{\chi}+(1+\hat\chi)\left[s\hat{T}+\frac{s^2}{2}\left(\hat{T}^2-\hat{\tau}_2\right)+\ldots\right]
\end{eqnarray}
Note that the first term is independent of $s$ and the remainder is projected onto
the non-hole part by $1+\hat\chi$.
\begin{equation}
\ln\la\chi_s\ra=\ln\la\chi\ra+s\tau_1+\frac{s^2}{2}\hat{\tau}_2+\ldots
\end{equation}
Note that higher terms in these expansions involve products of the various $\hat{\tau}_k$, and for $\hat{\chi}_s$,
powers of $\hat{T}$ as well.  The cumulants can similarly be expanded, but it turns out we will need only the leading
$s=0$ term, that is, replacing $\hat{\chi}_s$ by $\hat{\chi}$.

We need to establish the order of magnitude of the various quantities in the expansion.  First we
note that $\ln\la\chi\ra$ is of order $h$ (the hole size). The powers of $\hat{T}$ and
their correlations are of order unity.
Each factor of $\hat\chi$ in a correlation will result in a
factor of $h$ unless there is a high probability for orbits entering
the hole to return there.  In the case of a billiard in which a short periodic
orbit lies on the hole, an orbit leaving the hole needs to be in a precisely specified
direction in order to return once in a small number of collisions,
but after this it will return with high conditional probability.  Thus
correlations with two or more $\hat\chi$ terms should be of order $h^2$.
In non-generic billiards, (for example if the boundary contains an arc of a circle centred
on the hole that reflects orbits in many directions back to it) or for some non-billiard systems,
no fixing of direction is necessary and all correlations could be of order $h$.

More generally, the first order at which periodic orbits contribute depends on the dimension of
the system, the size of the hole in each dimension and the structure of the periodic
orbits (eg whether isolated).  Typical low dimensional chaotic systems have periodic
orbits that are dense, so all holes will cover sufficiently long periodic orbits.  However the
probability of following a long orbit for a whole period is very small, related to the exponential
of the Lyapunov exponent times the period.  Hence the term ``short periodic orbit'' means that
the set of orbits close to them have a probability of return large enough to make a measurable
impact on the escape rate. Previous work~\cite{BP01} has treated holes on short periodic orbits,
but limited to first order in $h$.

We note immediately that if we can assume the correlations of $\hat{\chi}$ are
small, the escape rate $\gamma$ given by the solution of $g(s)=0$ reduces simply to
$\gamma=-\ln\la\chi\ra/\tau_1$, which is the hole size divided by the average
collision time to leading order, as expected. It also means that $s$ is of order $h$.
Now, given our assumptions, we know the order of each of the quantities and we can
proceed to develop an expansion in the single parameter $h$.

We expand $\hat{\chi}_s$ into orders in $h$:
$\hat{\chi}_s=\hat{\chi}^{(1)}_s+\hat{\chi}^{(2)}_s+\ldots$ with
\begin{eqnarray}
\hat{\chi}^{(1)}_s&=&\hat{\chi}+s\hat{T}(1+\hat{\chi})\nonumber\\
\hat{\chi}^{(2)}_s&=&\frac{s^2}{2}(\hat{T}^2-\hat{\tau}_2)(1+\hat{\chi})
\end{eqnarray}
Note that while $s\hat{T}\hat{\chi}$ in $\hat{\chi}^{(1)}$ appears to be of higher order, it needs to
be included so that $\langle\hat{\chi}^{(k)}_s\rangle=0$ at each order; this is essential for convergence of the correlation sums.  
$g(s)$ now splits into terms based on their order in $h$, taking care that in general, $\hat{\chi}$ behaves differently to $s$: $g(s)=g^{(1)}(s)+g^{(2)}(s)+\ldots$ with
\begin{eqnarray}
g^{(1)}(s)&=&\ln\la\chi\ra+s\tau_1\\
g^{(2)}(s)&=&\frac{s^2}{2}\hat{\tau}_2+\sum_{0<j}\la\hat{\chi}^{(1)^0}_s\hat{\chi}^{(1)^j}_s\ra+Q_3(0)+Q_4(0)+\ldots \nonumber
\end{eqnarray}
Higher order contributions are more complicated, and most significantly, all involve an infinite series of correlations.  Far from a short periodic orbit, we can hope that $k$-order correlations are of order $h^k$, leading to a more tractable expansion,
with higher order cumulants relegated to higher orders in $s$.

Finally the solution of $g(s)=0$ gives the escape rate $\gamma$.
Writing (in orders of $h$)
$\gamma=\gamma^{(1)}+\gamma^{(2)}+\ldots$, expand $g(\gamma)$ in a
Taylor series about $\gamma^{(1)}$. Thus we have
\begin{equation}\label{e:g1}
\gamma^{(1)}=-\frac{\ln\la\chi\ra}{\tau_1},
\qquad \gamma^{(2)}=-\frac{g^{(2)}(\gamma^{(1)})}{\tau_1}
\end{equation}
This confirms the leading order behavior that the escape rate is the hole
size divided by the average collision time, and observe that the result
involves correlations of a quantity
\begin{equation}
\hat{\chi}^{(1)}_{s^{(1)}}=\hat{\chi}-\frac{\ln\la\chi\ra}{\tau_1}\hat{T}(1+\hat{\chi})
\end{equation}
which to leading order is equivalent to
\begin{equation}
u\equiv\hat{\chi}+\frac{h}{\la T\ra}\hat{T}(1+\hat{\chi})
\end{equation}
Thus we arrive at our first main result, giving an exact expansion for the
escape rate in powers of the hole size,
\begin{eqnarray}
\gamma&=&-\frac{1}{\tau_1}\left[\ln(1-h)+\frac{\hat\tau_2}{2}\left(\frac{\ln(1-h)}{\tau_1}\right)^2
+\sum_{j=0}^\infty\la u^0u^j\ra\right.\nonumber\\
&&\left.+\sum_{n=3}^\infty Q_n(0)+\ldots\right]\label{e:full}
\end{eqnarray}
where the higher cumulant terms are ignored for the purposes of the simulations
below and in general if the hole(s) avoid short periodic orbits.
\revision{If the hole(s) lie on short periodic orbits, relevant terms in the
higher cumulants would need to be summed explicitly.  This is beyond the scope
of this Letter, but should reproduce and generalise relevant equations in
Ref.~\cite{BP01}.}

Let us now compare the escape rates of a billiard with one hole of size $a$
and corresponding $\chi_A$, one (disjoint) hole of size $b$ and corresponding
$\chi_B$, and both holes, total size $a+b$ and corresponding
$\chi_{AB}=\chi_A+\chi_B-1$.  We compute (up to second order in the hole size)
\begin{eqnarray}
\gamma^{(1)}_{AB}&=&\gamma^{(1)}_A+\gamma^{(1)}_B+\frac{ab}{\la T\ra}-
\frac{b\la T\hat{\chi}_A\ra}
{\la T\ra^2}-\frac{a\la T\hat{\chi}_B\ra}{\la T\ra^2}+\ldots \nonumber\\
\gamma^{(2)}_{AB}&=&\gamma^{(2)}_A+\gamma^{(2)}_B-
\frac{ab\hat{\tau}_2}{\la T\ra^3}\\\nonumber
&&-\frac{1}{\la T\ra}\left\{\sum_{0<j}
\left[\la u^{0}_Au^{j}_B\ra+\la u^{0}_Bu^{j}_A\ra\right]\right.\\\nonumber
&&\left.+\sum_{n=3}^{\infty}\left[Q_{nAB}(0)-Q_{nA}(0)-Q_{nB}(0)\right]\right\}+\ldots
\end{eqnarray}
where $u$ for a single hole ($A$ or $B$) is defined above.  Note that no subscript need be given for $\hat{T}$ or $\hat{\tau}_2$ since to leading order they are equivalent to the ones of the closed system, namely $\hat{T}=T-\la T\ra$, $\hat{\tau}_2=\la\hat{T}^2\ra$. Putting this all together we find a remarkably simple relation, our second main result
\begin{eqnarray}\nonumber
\gamma_{AB}=\gamma_A+\gamma_B
-\frac{1}{\la T\ra}\left\{\sum_{j=-\infty}^\infty\la u^0_Au^j_B \ra\right.\\
\left.+\sum_{n=3}^{\infty}\left[Q_{nAB}(0)-Q_{nA}(0)-Q_{nB}(0)\right]\right\}+\ldots\label{e:ones}
\end{eqnarray}
The higher correlations all contain mixtures of $\hat\chi_A$ and $\hat\chi_B$,
so we expect they contribute at second order only if both holes lie on the
same short periodic orbit.
\revision{In Ref.~\cite{BP01} the difference $\gamma_{AB}-\gamma_A-\gamma_B$
is termed the ``interaction'' between the holes, and described approximately in terms of ``shadows''
cast by one hole on another, that is, overlap between one hole and the image of another under
the dynamics.  Here we give a precise formulation in terms of correlation
functions, which could also be extended to the three and multi-hole interactions described
there.  Clearly the strongest interactions exist when there are large shadow effects at short times.
Whether this is the case or not, the full (long time) correlation functions permit a
precise calculation.}

\begin{figure}
\centering\scalebox{0.5}{\includegraphics{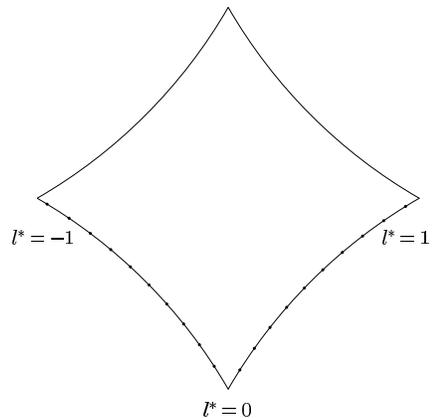}}
\caption{The billiard with $R=1/\sqrt{3}$ showing holes of size $R/20$ at
positions used in Fig.~2, and the $l^*$ coordinate used in the later figures.
\label{fig:bill}}
\end{figure}

Finally, we test the above formulas with numerical simulations.
We consider a ``diamond'' billiard, bounded by four overlapping disks of
radius $R$, centered at the corners of the unit square (Fig.~1).
When $1/2<R<1/\sqrt{2}$ this has strong chaotic properties probably
including exponential decay of multiple correlations~\cite{CD00,CM07}.
In the case $R=1/2$ the circles touch tangentially leading to the
possibility of long sequences of collisions near the corners (hence
weaker chaotic properties), and the limit $R\to 1/\sqrt{2}$ is the
integrable (regular) square.  Here we consider $R=1/\sqrt{3}$ which
leads to simple exact formulas for the perimeter
$|\partial D|=2\pi/3\sqrt{3}$, area $|D|=1-\sqrt{4/\sqrt{3}-1}-\pi/9$
and mean free path $\la T\ra=\frac{\pi|D|}{|\partial D|}$.  The latter formula
is general for two dimensional billiards~\cite{San02}.  The hole
position coordinate $l^*$ is defined so that a curved side corresponds to a
unit interval.

For the numerical simulation, a trajectory of $10^8$ collisions is
simulated and stored.  The position on the boundary is binned according
to a partition of the arcs into pieces of fixed length, giving
a single sequence of integers, from which escape rates of open systems
with holes given by any desired combinations of the pieces can be
calculated simultaneously. The correlation functions are also calculated
from the time series; for the infinite sums only a few (roughly ten) terms
need to be retained, due to the exponential decay of correlations for this system.

The choice of the size of the hole affects the numerical tests
in that for large holes higher order ($h^3$) terms become significant, while
for small holes the $h^2$ terms may be smaller than the errors in the
correlation statistics. 

\begin{figure}
\centering\scalebox{0.5}{\includegraphics{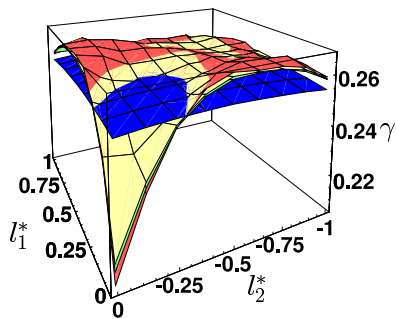}}
\caption{(Color online; monochrome versions have from light to dark: yellow,
green, red, blue) The escape rate $\gamma$ (red),
as a function of the position of two holes of size $R/20$, together with
the first (blue) and second (green) order approximations in Eq.~\ref{e:full}
and using one-hole escape rates (yellow) as in Eq.~\ref{e:ones}.
Note that the 3D graphics format does not conceal
large fluctuations, for example the RMS deviations from the directly calculated
escape rate are 0.00848 (first order), 0.00193 (second order)
and 0.00190 (second order using one-hole escape rates). \label{fig:col}}
\end{figure}

\begin{figure}
\centering\scalebox{0.65}{\includegraphics{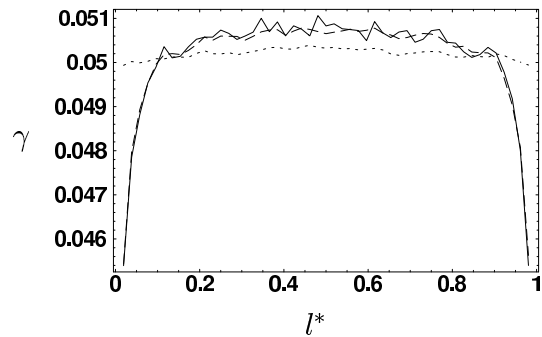}}
\caption{The escape rate $\gamma$ (solid) as a function of the
position of one hole of size $R/50$ at position $l^*$, together
with the first (dotted) and second (dashed) order approximations
in Eq.~\ref{e:full}.} 
\end{figure}

\begin{figure}
\centering\scalebox{0.6}{\includegraphics{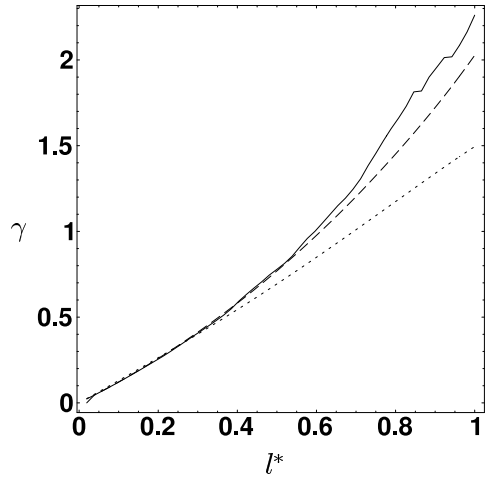}}
\caption{As for Fig.~3, except that the hole is of variable size, extending
from zero to $l^*$.}
\end{figure}

The results of the numerical simulations for some hole sizes and
positions are shown in Figs~2-4.  Clearly the second
order approximations are better than the first order approximation,
especially when both holes are near the same corner, which leads to a
lower escape rate.  Near a corner, we expect that some orbits have
several collisions in the holes, and hence the higher order correlation
functions are more important.  \revision{In Fig.~3 we note that fluctuations
in the escape rate as a function of hole position are visible; compare
with Ref.~\cite{STN02}.  Fig.~4 gives useful information about the validity
of the expansion truncated at the first or second term even when the
hole is no longer small.}

The numerical simulation thus confirm the formulas to second order,
which is already sufficient to determine detailed information about the
internal dynamics from observing escape rates as a function of hole position.
Computation of the necessary higher cumulants where holes are on short
periodic orbits will require a deeper analysis of the role of the latter.
Another important but only partially understood problem is that of general systems
with power law escape, such as stadium-type chaotic billiards, and systems with
mixed dynamics, where chaotic and regular behavior coexist.
We are confident that our approach (suitably modified) will work and
lead to general and useful formulas as well, as suggested by
similarities between results in the strongly chaotic systems considered
here and in regular systems~\cite{BD05}.

L.B. was partially supported by NSF grant DMS-0140165.

\end{document}